\documentclass[11pt]{amsart}
\usepackage{amsmath, amssymb}
\usepackage{upgreek}
\usepackage{xcolor, url}
\usepackage[all]{xy}
\usepackage{graphicx}
\usepackage{bm}
\usepackage{enumerate}
\theoremstyle{plain}

\newcommand{\td}{\text{d}}
\theoremstyle{definition}

\usepackage{subfig, tikz}
\usetikzlibrary{decorations.pathmorphing}
\usepackage{graphics}
\usepackage{cite}

\usepackage{hyperref} 
\hypersetup{colorlinks=true, linkcolor=blue, citecolor = blue, urlcolor=blue, filecolor=magenta}

\usepackage{amsmath}

\allowdisplaybreaks

\begin{document}
\title[Ovcharenko-Podolsky Einstein-Maxwell gravitational instantons]{Ovcharenko-Podolsky Einstein-Maxwell gravitational instantons}

\author{Hari K. Kunduri}
\address{Department of Mathematics and Statistics and Department of Physics and Astronomy\\
		McMaster University\\
		Hamilton, ON Canada}
	\email{kundurih@mcmaster.ca}

\begin{abstract}
We present an explicit three-parameter family of toric Einstein-Maxwell gravitational instantons. These are complete Riemannian manifolds with vanishing scalar curvature which satisfy the Riemannian Einstein-Maxwell equations. We argue that the metric can be extended to a space with two asymptotic ends diffeomorphic to $\mathbb{H}^2 \times S^2$ with a non-collapsing two-cycle (bolt) in the bulk.  The global metric is produced by suitably extending an analytic continuation of a local family of Lorentzian black hole geometries constructed by Ovcharenko-Podolsky.
\end{abstract}

\maketitle
\section{Introduction}
A gravitational instanton is a complete four dimensional Riemannian manifold $(M,g)$ satisfying the vacuum Einstein field equations with prescribed curvature decay.  They are characterized by the geometry and topology of their asymptotic ends. Of particular interest are asympotically locally flat (ALF) instantons with an end diffeomorphic to $\mathbb{R} \times L$ where $L$ is an $S^1$ bundle over $S^2$ (with cubic volume growth) and asymptotically locally Euclidean (ALE) instantons with an end diffeomorphic to a discrete quotient of $\mathbb{R}^4$ equipped with the Euclidean metric (with quartic volume growth). For precise definitions of these asymptotic classes see e.g. \cite{Biquard:2021gwj} and\cite{Alaee:2026wty}. An important subset of ALF instantons are asymptotically flat (AF) for which the $L = S^1 \times S^2$ is the trivial bundle.  Explicit examples of gravitational instantons were classically constructed by analytically continuing Lorentzian black hole metrics. For example, the Schwarzschild and Kerr black hole solutions can be used to construct AF gravitational instantons on $\mathbb{R}^2 \times S^2$; similar constructions yield ALF instantons on $\mathbb{R}^4$ and $\mathbb{CP}^2 \setminus \{p\}$ starting from the Taub-NUT and Taub-Bolt solutions respectively. The well-known Eguchi-Hanson gravitational instanton on $T^* S^2$ provides the classic example of an ALE instanton. There has been recent renewed interest in the classification of gravitational instantons since Chen and Teo produced an explicit example of a two-parameter family of AF gravitational instanton on $\mathbb{CP}^2 \setminus S^1$~\cite{Chen:2011tc}. 

In the special case in which $(M,g)$ admits a toric action, the vacuum Einstein equations reduce to a harmonic map from $\mathbb{R}^3 \setminus \Gamma \to \mathbb{H}^2$ where $\Gamma$ may be chosen to be the $z-$axis \cite{Kunduri:2021xiv} . In this case, a set of invariants which characterize the torus action (referred to as `rod structures') can be defined, and existence and uniqueness results for AF \cite{Kunduri:2021xiv} and ALF and ALE \cite{Kunduri:2026xvc} gravitational instantons can be proved. These results establish that for a given rod structure and a small set of moduli which parameterize the asymptotic behaviour of the metric, there exists a unique solution which is smooth everywhere apart from possible conical singularities along the degeneration set of the torus action.  Most remarkably, Li and Sun have proven that a subset of these abstractly constructed solutions can be made regular in the AF setting, producing an infinite set of such solutions \cite{Li:2026zzn}. Teo has subsequently recently produced an explicit new example of an AF instanton of this type \cite{Teo:2026kej}. Classification results have also been obtained under various other natural geometric restrictions, e.g., for ALF \cite{Biquard:2021gwj} and ALE \cite{Araneda:2025uqo}  Hermitian gravitational instantons, conformally K\"ahler \cite{Li:2023jlq}, and $S^1-$invariant instantons \cite{Aksteiner:2023djq}. 

It is natural to extend the definition of gravitational instanton to other positive signature metrics satisfying a natural curvature equation. Thus, one may consider gravitational instantons which are Einstein, i.e. $\text{Ric}(g) = \lambda g$ for some constant $\lambda \neq 0$; for $\lambda >0$ the classic example is the compact Page instnaton on $\mathbb{CP}^2 \# \overline{\mathbb{CP}}^2$ \cite{Page:1978vqj} or the Euclidean-Schwarzschild-AdS metric with $\lambda <0$. For a recent classification of toric self-dual instein gravitational instantons, see \cite{Araneda:2026wxn,Araneda:2026uag}.  Another natural set are instantons which satisfy the coupled Riemannian Einstein-Maxwell equations 
\begin{equation}\label{E-M}
    R_{ab} = \mathcal{F}_{ac}\mathcal{F}_b^{~c} -(\star\mathcal{F})_{ac}(\star\mathcal{F})_b^{~c} , \qquad d\mathcal{F} =0, \qquad d \star \mathcal{F} =0.
\end{equation} Note that it follows from taking the trace of \eqref{E-M} that $(M,g)$ has vanishing scalar curvature.  Examples may be readily constructed, again by suitable analytic continuations of Lorentzian metrics (see, e.g. \cite{Mellor:1989wc} and more recently \cite{Tod:2024msz}). A particularly remarkable new example is the three-parameter family of AF Einstein-Maxwell instantons constructed by Araneda-Dunajski \cite{Araneda:2025lak} which generalize the Chen-Teo family. 

The Riemannian Einstein-Maxwell equations allow for a much richer set of asymptotic behaviour, in particular, the possibility of multiple asymptotic ends. Recall the Cheager-Gromoll splitting theorem \cite{Cheeger:1971nn}, which implies that a manifold with positive semi-definite Ricci tensor $\text{Ric} \geq 0$ can have at most two ends, and moreover, if it does have two ends, then it splits isometrically, i.e. it has topology $M = \mathbb{R} \times N$ where $N$ is compact and $g$ is a product metric. If $(M,g)$ is further assumed to be Ricci flat, then $N$ must be a quotient of $\mathbb{R}^3$ with a locally flat metric.  This argument forbids vacuum ALF/AF  and ALE gravitational instantons with multiple ends.  Of course, this argument fails in the Einstein-Maxwell case as the Ricci tensor need not be definite (indeed it cannot be positive definite, since it is scalar flat). A familiar example would be the set of positive signature scalar flat geometries which can be produced from the extreme Reissner-Nordstr\"om black hole. These will have asymptotically cylindrical ends along with an AF one. 

A simple model solution of the system \eqref{E-M} with two asymptotic ends is $(\mathbb{R} \times S^1 \times S^2,b)$ where
\begin{equation}\label{model}
    b = (1 + w^2 ) \td \tau^2 + \frac{\td w^2}{1 + w^2} + g_{\mathbb{S}^2}. \qquad \mathcal{F} = \text{vol}_{S^2},
\end{equation} and $\tau \sim \tau + 2\pi, w \in \mathbb{R}$. The $(\tau,w)$ part of the metric is the standard hyperbolic metric on the cylinder. The geometry admits a torus action generated by $\partial_\tau$ and the rotational Killing field on the round $S^2$. Note that $\partial_\tau$ has no fixed points,and there are in particular no 2 cycle (bolts) in the interior. The isometry subgroup of $b$ connected to the identity is simply $U(1) \times SO(3)$ (there is also a $\mathbb{Z}_2$ which acts by $w \to -w$). If $\tau$ takes values on the real line, then one recovers the full hyperbolic plane and there is only one asymptotic end. 

This simple geometry can of course be thought of as a Riemannian analogue of AdS$_2 \times SS^2$ with an identification on the time coordinate. The purpose of this note is to construct a scalar-flat gravitational instanton solution of \eqref{E-M} which we will argue behaves similarly to \eqref{model} in its asymptotic regions, but also contains a 2-cycle in the interior. The local solution is produced by analytically continuing a remarkable stationary and axisymmetric Lorentzian black hole solution recently constructed by Ovcharenko and Podolsky~\cite{Podolsky:2025tle, Ovcharenko:2025cpm}. The novel solution physically describes a Kerr black holes immersed in an external magnetic field. The Maxwell field is non-laigned with the two principal null directions of the Weyl tensor in contrast to other algebraically special solutions. Section 2 presents the local metric and analyzes the global behaviour and focuses on a simpler `static' subset which illustrates the main properties of this class of instantons. \\
\par\noindent
{\it Note added}: while this work was nearing completion, a global analysis of the Loretnzian signature black hole metrics discovered in \cite{Podolsky:2025tle} appeared which complements the present work~\cite{Ovcharenko:2026ooh}. We emphasize that although the local metrics appear superficially similar, the difference in signatrue creates significant differences. 

\subsection*{Acknowledgments} We thank James Lucietti for many useful discussions and Marcus Khuri for pointing out the relevance of the splitting theorem, as well as the Simons Center for Geometry and Physics for hospitality while this work was initiated. The work is supported by NSERC Discovery Grant RGPIN-2025-0602.

\section{The Ovcharenko-Podolsky instanton}
\subsection{The local family of metrics}

The following  family of metrics has vanishing scalar curvature: 
\begin{equation}\label{gext}
\begin{aligned}
    g = \frac{1}{A^2} \left[\frac{B(r)}{\Sigma} \left(\td \tau - \alpha \sin^2\theta \td \phi\right)^2 + \frac{C(\theta)}{\Sigma} \sin^2\theta (\alpha \td \tau + (r^2 - \alpha^2) \td \phi)^2 + \Sigma \left(\frac{\td r^2}{B(r)} + \frac{\td \theta^2}{C(\theta)} \right)\right]
    \end{aligned}
\end{equation} where 
\begin{align}
   \Sigma:= \Sigma(r,\theta) &= r^2 - \alpha^2 \cos\theta^2, \qquad C(\theta) = 1 + \beta^2 \left(\frac{m^2 I_2}{I_1^2} + \alpha^2 \right) \cos^2\theta, \\
    B(r) & = (1 + \beta^2 r^2) \Delta(r), \quad \Delta(r) = \left(1 - \beta^2 m^2 \frac{I_2}{I_1^2}\right) r^2 - \frac{2m I_2}{I_1} r - \alpha^2 \\
    A &= \left[(1 + \beta^2 r^2) - \beta^2 \Delta(r) \cos^2\theta \right]^{1/2}
\end{align} with
\begin{equation}
    I_1 = 1 + \frac{\beta^2 \alpha^2}{2}, \qquad I_2 = 1 + \beta^2 \alpha^2.
\end{equation} The solution is parametrized by three real parameters $(\alpha, \beta, m)$ and local coordinates $(\tau,r,\theta,\phi)$. The ranges of these parameters and coordinates will be determined below to ensure that the metric above extends to a smooth global metric.  The local family of metrics is obtained from the family of Lorentizian black hole solutions constructed recently in \cite[Eq 2.1]{Podolsky:2025tle} by sending their parameter $a \to i \alpha$ and passing to Euclidean time $t \to i \tau$. The Maxwell two-form is given by 
\begin{equation}
    \mathcal{F} = \td \mathcal{A}, \qquad \mathcal{A} = \frac{1}{2\beta}\left[ \partial_r A \frac{\alpha \td \tau + (r^2 - \alpha^2) \td \phi}{r - \alpha \cos\theta} + \frac{\partial_\theta A}{\sin\theta} \frac{\td \tau - \alpha \sin^2\theta \td \phi}{r - \alpha \sin\theta} - (A - 1) \td \phi \right]
\end{equation} It can be verified that $(g,\mathcal{F})$ constitute a local solution to the system \eqref{E-M}. The solution is invariant under the torus action generated by the two commuting Killing vector fields $\partial_\tau, \partial_\phi$. 

Let $g_K$ denote the restriction of the metric to the Killing fields generating the torus action. Since
\begin{equation}
   \rho:= \sqrt{\det g_K} =  \frac{\sqrt{B(r)C(\theta)} \sin\theta}{A^2}
\end{equation} we see that at least one of linear combination of Killing fields will degenerate on the zero set of $\rho$. This set consists of the points $\theta= 0, \pi$ and any zeroes of $B(r)$ or $C(\theta)$.   The solution $(g,F)$ admits an isometric torus action generated by the commuting Killing vector fields $\partial_\tau, \partial_\phi$. Therefore the solution falls into the class of Weyl metrics. We denote by $g_2$ the two-dimensional metric induced on the space of orbits. As in the vacuum, it can be shown that the field equations \eqref{E-M} imply that the function $\rho$ is harmonic with respect to the orbit space metric
\begin{equation}
    g_2 = \frac{1}{A^2} \left[ \Sigma \left(\frac{\td r^2}{B(r)} + \frac{\td \theta^2}{C(\theta)} \right)\right] ,
\end{equation} so that $\Delta_2 \rho =0$. Its harmonic conjugate $z$, defined up to an integration constant by $\td z = -\star_2 \td \rho$, is given by 
\begin{equation}
    z = \frac{(1 + \alpha^2 \beta^2)((2 + \alpha^2 \beta^2)r - 2m)(2 + \alpha^2 \beta^2 + 2 \beta^2 m r)\cos\theta }{(2 + \alpha^2 \beta^2)^2 A^2}
\end{equation} The orbit space metric is then
\begin{equation}
    g_2 = e^{2\nu} (\td \rho^2 + \td z^2), \qquad  e^{2\nu} = \frac{(2 + \alpha^2\beta^2)^6\Sigma A^2}{(1 + \alpha^2\beta^2)^2 U_1 U_2} 
\end{equation} with
\begin{align*}
    U_1  &= (2 + (2m r + \alpha^2)\beta^2)^2 + \beta^2((2 + \alpha^2\beta^2)r - 2m)^2 \cos^2\theta \\
    U_2  &= (2 + \alpha^2\beta^2)^2((2 + \alpha^2\beta^2)r - 2m)^2 - ((2 + \alpha^2\beta^2)^2\alpha^2 +4m^2(1 + \alpha^2\beta^2)) \\ & \times (2 + (2mr + \alpha^2)\beta^2)^2 \cos^2\theta
\end{align*}

We assume that $m, \alpha, \beta \geq 0$. The coordinates $(\theta,\phi)$ are simply related to standard spherical coordinates on $S^2$ with inhomogeneous metric. The constants $I_1, I_2$ are clearly positive and it follows that $C(\theta)>0$. 
The zeroes of $B(r)$ are determined by the zeroes of $\Delta(r)$, which are  
\begin{equation}
    r_\pm = \frac{ m I_2 \pm  \sqrt{ m^2 I_2 + \alpha^2 I_1^2}}{I_1^2 - \beta^2 m^2 I_2} I_1
\end{equation} To ensure that $r_+ > 0$, we impose the bound
\begin{equation}\label{mbound}
    m < \frac{2 + \alpha^2 \beta^2}{2 \beta \sqrt{1 + \alpha^2 \beta^2}}
 \end{equation}  This condition ensures that the quadratic $\Delta(r)$ is convex, and in particular positive for $r > r_+>0$. Using \eqref{mbound} we find that $r_- < 0$. Thus $r_- < 0 < r_+$. Furthermore, as a consequence of \eqref{mbound} we find that 
 \begin{equation}
     r_+ > \frac{\alpha}{\left(1 - \frac{\beta^2 m^2 I_2}{I_1^2}\right)} > \alpha.
 \end{equation} This ensures that $\Sigma > 0$ for $r \geq r_+$. Finally, note that \eqref{mbound} implies the following inequalities:
 \begin{equation}\label{mbound2}
 m < \frac{2 + \alpha^2 \beta^2}{2 \beta}, \qquad m <\frac{2 + \alpha^2 \beta^2}{2 \beta^2 \alpha} .
\end{equation} The metric \eqref{gext} is now seen to be positive definite and non-singular away from the zeroes of $\rho$, where the torus action degenerates. In particular, using \eqref{mbound}, the conformal factor $A^2$ is manifestly positive since
\begin{equation}
    A^2 = 1 + \beta^2 (r^2 + \alpha ^2 x^2) + \left[ \frac{4\beta^2 (1 + \alpha^2 \beta^2) m r}{2 + \alpha^2 \beta^2} + \left(1 - \frac{4\beta^2 (1 + \alpha^2\beta^2)m^2}{(2 + \alpha^2\beta^2)^2}\right)r^2\right]x^2 > 0
\end{equation} where we have set $x = \cos\theta$ with $x \in [-1,1]$. 

We now discuss the axis set, which consists of three components, which we will refer to as `rods' following the standard nomenclature. The torus parameterized by $(\tau,\phi)$ degenerates along : (1) a semi-infinite rod $\theta =0, r > r_+$, a finite rod $r = r_+, \theta \in (0,\pi)$, and a semi-infinite rod $\theta = \pi, r> r_+$. The same Killing field degenerates on the two semi-infinite rods. The generator of $2\pi-$periodic closed orbits is given by
\begin{equation}
    \ell_1 =b_1 \frac{\partial}{\partial \phi}, \qquad b_1:=\frac{1}{1 + \alpha^2 \beta^2 + \gamma^2} 
\end{equation} where we have defined
\begin{equation}
    \gamma: = \frac{2\beta m \sqrt{1 + \alpha^2\beta^2}}{2 + \alpha^2\beta^2} < 1
\end{equation} with inequality following from \eqref{mbound}. There is also a finite rod $r = r_+, 0 < \theta < \pi$ for which the Killing vector field 
\begin{equation}
    \ell_2 = b_2\left( \frac{\partial}{\partial \tau} - \frac{\alpha}{r_+^2 - \alpha^2} \frac{\partial}{\partial \phi} \right)
\end{equation}  degenrates smoothly (that is $\ell_2$ generates $2\pi-$periodic orbits).  Here $b_2$ is a constant although its expression is fairly complicated. It reduces to the usual Kerr surface gravity for $\beta=0$. Explicitly,
\begin{equation}
    \begin{aligned}
        b_2 & = \frac{L_2}{L_1} \\
        L_1 & = k_1 (\alpha ^8 \beta ^8+16 \beta ^2
   m \left(16 \alpha ^2
   (\beta ^2 m
   (k+m)+\beta ^4 m^4+2\right)  \\ & +\beta ^2 k
   m^3+4 \alpha ^4 \beta ^2
   \left(\beta ^2 m (k+4
   m)+6\right)) +4 \alpha
   ^6 \beta ^6 \left(\beta ^2
   m^2+2\right)+16) \\
   L_2 & = 4 (2 + \alpha^2\beta^2) m \left[(2 + \alpha^2 \beta^2)^2(2 + 3 \alpha^2\beta^2)m - 4\alpha^2\beta^4 m^3(1 + \alpha^2\beta^2) + (2 + \alpha^2\beta^2)^2 k_1\right], \\
   k_1 & = \left[\alpha^2 (2 + \alpha^2\beta^2)^2 + 4 (1 + \alpha^2\beta^2) m^2\right]^{1/2}.
    \end{aligned}
\end{equation} where $\eqref{mbound}$ ensures $b_2>0$. Since $(\ell_1, \ell_2)$ generate independent $2\pi$ periodic rotations, we must impose the following identifications in the $(\tau,\phi)$ plane: 
\begin{equation}
    (\tau,\phi) \sim (\tau, \phi + 2\pi b_1), \qquad (\tau, \phi) \sim (\tau + 2\pi b_2, \phi - 2\pi\Omega_H b_2)
\end{equation} where we have defined 
\begin{equation}
    \Omega_H = \frac{\alpha}{r_+^2 - \alpha^2} \geq 0.
\end{equation} Note that $\ell_2$ is proportional to $\partial_\tau$ if and only if $\alpha =0$; in this case, the $\tau$ coordinate must be identified with period $2\pi b_2$: in this case $\partial_\tau$ is hypersurface orthogonal and the geometry is `static' in a Riemannian sense. Otherwise, the $\tau$ coordinate does not generate closed orbits. 

\subsection{Asymptotic geometry}
For $\beta \equiv 0$, the metric \eqref{gext} reduces to that of the Kerr gravitational instanton \cite[Appendix B]{Kunduri:2021xiv}, which is asymptotically flat (AF) in the sense of a gravitational instanton. This means that the Riemannian manifold $(M,g)$ has an asymptotic end diffeomorphic to $\mathbb{S}^1 \times \mathbb{R}^3$ equipped with the flat metric (note that the $S^1$ does not generically have bounded size). For $\beta > 0$ howwver, we will demonstrate that this is not the case, and we can extend the metric beyond $r \to \infty$.  The leading order behaviour of the metric components in the limit $r \to \infty$ is given by 
\begin{equation}
\begin{aligned}
    g &\to \frac{1}{U(\theta)} \left[ (1- \gamma^2)(\td \tau - a \sin^2\theta \td \phi)^2  + \frac{\td r^2}{\beta^4 (1-\gamma^2) r^4} + \frac{\td \theta^2}{\beta^2 (1 + (\alpha^2\beta^2 + \gamma^2)\cos^2\theta)} \right. \\ & \left. + \frac{\sin^2\theta \td \phi^2}{\beta^2} (1 + (\alpha^2 \beta^2 + \gamma^2)\cos^2\theta)\right]
    \end{aligned}
\end{equation} where
\begin{equation}
    U(\theta) = \sin^2\theta + \gamma^2 \cos^2\theta.
\end{equation} The volume form $\td \text{vol}(g)\sim r^{-2} \td r \wedge \wedge \td \tau \wedge \td \theta \wedge \td \phi$ which implies that the volume of this region is finite as $r\to \infty$, and certainly does not diverge to cubic order, as would be the case in the AF setting. Define a new coordinate \begin{equation}\label{ucoord} u := -1/r 
\end{equation} so that $\td u^2 = \td r^2/ r^4$. Observe that the volume growth is clearly linear in the coordinate $u$. This implies that the region $r_+ < r < \infty$, which maps to $u_+:=-1/r_{+} < u < 0$, has finite volume. It is now clear that the metric  be extended through $u = 0$ to a new region with $u>0$. 

Potential singularities in the metric can occur at zeroes in the functions $A, \Sigma$, and $B$ when $u > 0$. We examine these possibilities in turn. By direct computation of the Kretchmann scalar $R_{abcd} R^{abcd}$, it can be checked that the curvature tensor diverges at any zero of $\Sigma$. Since $\Sigma = u^{-2}(1 -  u^2 \alpha^2 \cos^2\theta)$, the singular set occurs at $u^2 = 1/(\alpha^2 \cos^2\theta) \geq 1/\alpha^2$. Therefore, requiring that $u < \alpha^{-1}$ will prevent the above curvature singularity from arising. 

Next we consider additional zeroes of $\Delta$ that arise in the extended region $u>0$. Observe that, since $r_- < 0 < r_+$, it follows $u_+ := -1/r_+ < 0$ and $u_- := -1/r_- > 0$. At $u= u_-$ , the torus area function $\rho=0$ and hence this set would constitute another finite axis rod. We will demonstrate, however, that this possibility does not occur. 

Now the conformal factor will diverge when $A^2 = A(r,\theta)^2$ vanishes. Let us assume that the blow up of the  conformal factor characterizes an asymptotic region, similar to the $y, x \to -1$ limit in the standard C-metric coordinates.  The conformal factor $A^2$ vanishes on the zero set of the quadratic in $u$. The discriminant of this quadratic is
\begin{equation}
    -4\beta^2 (2 + \alpha^2 \beta)^2((2 + \alpha^2\beta^2)^2 +  x^2 \left[\alpha^2 \beta^2(4 + 4\alpha^2 \beta^2 + \alpha^4 \beta^4) + 4 \beta^2 m^2(1 + \alpha^2\beta^2)\right] )(1 - x^2).
\end{equation} where recall $x = \cos\theta$. The discriminant is manifestly non-positive and vanishes if and only if $x = \pm 1$. Thus only one real root can appear with multiplicity 2. This unique double root is given by
\begin{equation}\label{ustar}
    u_* = \frac{2 m \beta^2}{2 + \alpha^2 \beta^2}.
\end{equation} A necessary condition for regularity is to ensure that $0 < u_* < u_- = -1/r_-$ and $u_* < 1/\alpha$. This guarantees that the conformal factor diverges before a zero of $\Delta$ and $\Sigma$ is reached. 

We can demonstrate that $u_* < 1/\alpha$ and $u_* < u_-$ for all $(m,\alpha, \beta)$ provided that the inequality \eqref{mbound} holds, as follows. Direct computation shows that positivity of $u_- - u_*$ requires
    \begin{equation}
    (2 + \alpha^2 \beta^2)^2 > 2 \beta^2 m \sqrt{(2 + \alpha^2 \beta^2)^2 a^2 + 4m^2 ( 1 + \alpha^2 \beta^2)} 
    \end{equation} Both sides of this inequality are positive and this inequality is equivalent to 
    $(2 + \alpha^2 \beta^2)^2 - 4 m^2 \beta^2 ( 1 + \alpha^2\beta^2) > 0$
    which is true as a consequence of \eqref{mbound}. Next we compute
    \begin{equation}
        \frac{1}{\alpha} - u_* = \frac{2 + \alpha^2\beta^2 - 2 \alpha m \beta^2}{\alpha(2 + \alpha^2\beta^2)} > \frac{\sqrt{1 + \alpha^2\beta^2} - \alpha \beta}{\alpha \sqrt{1 + \alpha^2\beta^2}} > 0.
    \end{equation}
    
The smooth hypersurface $u=0$ corresponds to the original region $r \to \infty$. 
The expansion of $A^2(r,\theta)$ about $u = u_*$ and $x = \cos\theta = 1$ (that is, $\theta =0$) is 
\begin{equation}\begin{aligned}
    A^2(u,\theta(x)) &= \frac{(1+ \alpha^2 B^2)(2 + \alpha^2 B^2)^2}{4 B^4 m^2} (u_*-u)^2 + \frac{(2 + \alpha^2 B^2)^2 + 4B^2 m^2}{2 B^2 m^2}(1-x)  \\ &+ \frac{(2 + \alpha^2 B^2)^3}{2 B^4 m^3}(u_*-u)(1 - x) + \ldots
    \end{aligned}
\end{equation} 
The local metric in the $(u,x,\tau, \phi)$ coordinate system clearly degenerates as $u = u_*, x - \pm 1$. Before introducing charts adapted to this region, we observe that by explicit computation using {\sc Mathematica}, the curvature remains finite in this region. In paritcular as $u \to u_*, \theta \to 0,\pi$ the curvature invariant is finite: 
\begin{equation}
    R_{abcd}R^{abcd} = \frac{8 B^4 (1 + \alpha^2 B^2)^2 ((2 + \alpha^2 B^2)^2 + 4 B^2 m^2)^4}{(2 + \alpha^2 B^2)^4((2 + \alpha^2 B^2)^2 - 4 B^4 m^2)^4}.
\end{equation} Observe that the inequality \eqref{mbound} ensures that the denominator of the above expression  does not vanish. Furthermore, the space is conformally flat,
\begin{equation}
    W_{abcd}W^{abcd} = 0
\end{equation} where $A^2=0$. Finally, consider $\det g_K$, which is a measure of the volume element on the $\mathbb{T}^2$ directions. For $u \neq u_*$ this vanishes at $x = \pm 1$ since these are axes; but when $u = u_*$, it diverges as $x \to \pm 1$ as $\det g_K \sim (1\pm x)^{-1}$. This suggests that the region where $A^2$ vanishes is a true asymptotic region, in the sense that there is unbounded volume growth in that region.  We shall now investigate this expectation more carefully. 

Consider first the region $u \to u_*, x \to 1$. Introduce coordinates $(w,y)$ by:
\begin{equation}
    w = \left[c_1^2 (u_* - u)^2 + 2 c_2^2 (1 - x) \right]^{-1/2}, \qquad  y = \left[ 1 + \frac{2 c_2^2 ( 1 - x)}{c_1^2 (u_* - u)^2} \right]^{-1/2}
\end{equation}
 where 
\begin{equation}
    c_1 = \frac{(2 + \alpha^2 B^2)\sqrt{1 + \alpha^2 B^2}}{2 B^2 m}, \qquad c_2 = \frac{\sqrt{(2 + \alpha^2 B^2)^2 + 4 B^2 m^2}}{2 B m}.
\end{equation} The inverse is given by
\begin{equation}
    u = u_* - \frac{y}{c_1 w}, \qquad x = 1 - \frac{(1-y^2)}{2 c_2^2 w^2}
\end{equation} Note that $w > 0$ and $w \to \infty$ if and only if $u \to u_*$ and $x \to 1$. We also have $0 < y \leq 1$. In particular on the axis $x = 1$ for any $u < u_*$, we have $y =1$ and $y \to 0^+$ as $u \to u_*$ and $x < 1$. We are mainly interested in using the $(w,y)$ chart for the region $u \in (0,u_*)$. The restriction $x \in [-1,1]$ translates to the requirement
\begin{equation}\label{condwy}
    \frac{1-y^2}{w^2} \leq 4 c_2^2.
\end{equation} We can then work out the expansion of the metric as we approach the asymptotic region $w \to \infty$. As expected, one finds
    \begin{equation}
    R_{abcd}R^{abcd} = \frac{8 B^4 (1 + \alpha^2 B^2)^2 ((2 + \alpha^2 B^2)^2 + 4 B^2 m^2)^4}{(2 + \alpha^2 B^2)^4((2 + \alpha^2 B^2)^2 - 4 B^4 m^2)^4} + O(w^{-2})
\end{equation} and $W_{abcd}W^{abcd} = O(w^{-4})$. Define the positive constant
\begin{equation}
    \lambda_1 = \frac{\beta^2(1 + \alpha^2\beta^2)((2 + \alpha^2\beta^2)^2 + 4 \beta^2 m^2)^2}{(2 + \alpha^2\beta^2)^2((2 + \alpha^2\beta^2)^2 - 4 \alpha^2\beta^4 m^2)}.
\end{equation} Thus as $w \to \infty$ the full metric has the asymptotic behaviour 
\begin{equation}\label{End1}
    \begin{aligned}
        g_{tt} &= \frac{(2 + \alpha^2 \beta^2)^2 \lambda_1}{4\beta^4(1 + \alpha^2\beta^2) m^2} w^2 \left(1 + O(w^{-1})\right), \qquad g_{ww} = \frac{1}{\lambda_1 w^2} \left(1 + O(w^{-1})\right), \\ 
        g_{wy} & = -\frac{16 \beta^2 m^2 y^2}{\sqrt{1 + \alpha^2\beta^2}((2 + \alpha^2 \beta^2)^2 + 4\beta^2 m^2) \lambda_1 w^2} \left(1 + O(w^{-1})\right) \\ g_{yy} &= \frac{1}{\lambda_1 (1 - y^2)} \left(1 + O(w^{-1})\right), \\
        g_{\phi\phi}  &=\frac{1 + \alpha^2\beta^2)((2 + \alpha^2\beta^2)^2 - 4 \alpha^2 \beta^4 m^2)}{\beta^2 ( 2 + \alpha^2 \beta^2)^2} (1-y^2) \left(1 + O(w^{-1})\right) \\
    & =    \frac{(1-y^2)}{\lambda_1}  \left(\frac{(1 + \alpha^2 \beta^2)((2 + \alpha^2\beta^2)^2 + 4 \beta^2 m^2)}{(2 + \alpha^2\beta^2)^2}\right)^2 \left(1 + O(w^{-1})\right) \\
        g_{t\phi} & = -\alpha (1-y^2) \left[1 + \frac{16 a^2 \beta^6(1 + \alpha^2 \beta^2) m^4}{(2 + \alpha^2 \beta^2)^4 - 4\alpha^2 \beta^4(2 + \alpha^2 \beta^2)^2 m^2} \right].
    \end{aligned}
\end{equation} The asymptotic geometry is not Einstein, but in the limit as $w \to \infty$, the Ricci curvature simplifies
\begin{equation}
    R^a_{~b} = \begin{pmatrix} -\lambda_1  + O(w^{-2}) & 0 & 0 & 0 \\
    0 & -\lambda_1 + O(w^{-2}) & O(w^{-1}) & 0 \\
    0 & O(w^{-3}) & \lambda_1 & 0 \\
    c + O(w^{-2}) & 0 & 0 &\lambda_1 + O(w^{-2}) \end{pmatrix}
\end{equation} where $c$ is a constant. In an appropriate basis of eigenvectors $\{ \partial_t - c/(2\lambda_1) \partial_\phi, \partial_r, \partial_\theta, \partial_\phi \}$, the Ricci endomorphism $R^a_{~b} \to \text{diag}(-\lambda_1, -\lambda_1, \lambda_1, \lambda_1)$ characteristic of the product of negative and a positive curved Einstein metric with equal radii of curvature. 

The asymptotic metric \eqref{End1} describes a twisted product of a locally hyperbolic metric $H^2$ and $D^2$ where $D^2$ is an open disc parameterized by $0 < y \leq 1$ with $y=1$ and $y=0$ corresponding respectively to the centre of the disc and its boundary. For $\alpha \neq 0$, the orbits of $\partial_\tau$ are not closed and thus in the asymptotic region, the $(w,\tau)$ coordinates cover hyperbolic plane minus a compact ball. For $\alpha=0$,  $\partial_\tau$ generates closed orbits and asymptotically $(w,\tau)$ coordinates cover the exterior region of a hyperbolic funnel $(0,\infty)\times S^1$. However, the geometry is clearly smooth through $y =0$. We may extend the range of $y$ to $y \in [-1,1]$. It can be checked that $y=-1$ is a coordinate singularity where $\partial_\phi$ smoothly degenerates. The result is that $(y,\phi)$ now parameterize an $S^2$. Note that the region $y \leq 0$ is not part of the original part of the manifold covered by the $(u, x)$ chart.  

Note that for large $w$, the Weyl coordinates have the form
\begin{align}
    \rho &= \frac{\sqrt{1 + \alpha^2 B^2}((2 + \alpha^2 B^2)^2 + 4 B^2 m^2) w \sqrt{1 - y^2}}{2 B^2 m (2 + \alpha^2 B^2)} + O(1) \\
    z & = -\frac{\sqrt{1 + \alpha^2 B^2}((2 + \alpha^2 B^2)^2 + 4 B^2 m^2) w y }{2B^2 m (2 + \alpha^2 B^2)} + O(1) 
\end{align} Thus the Weyl coordinate functions $(\rho,z)$ asymptotically behave similarly to those of the model \eqref{model} for which it is easily checked that $\rho = \sqrt{1 + w^2}\sqrt{1-y^2} \sim w \sqrt{1-y^2}$ and $z = wy$ where $y$ is the height function on the unit round $S^2$ metric. 

Note that at other point where the conformal factor $A^2$ diverges ($x \to -1$), $w$ takes a positive finite value and $y =0$. Thus this region is mapped to a point in this chart and the metric is still singular. We therefore introduce coordinates $(\hat{w}, \hat{y})$ adapted to $u\to u_*, x\to -1$ satisfying
\begin{equation}
    \hat{w} = \left[ c_1^2(u - u_*)^2 + 2 c_2^2 (1 + x)\right]^{-1/2}, \qquad \hat{y} = -\left[1 + \frac{2 c_2^2 (1+x)}{c_1^2 (u- u_*)^2}\right]^{-1/2},
\end{equation} with inverse 
\begin{equation}
    u = u_* + \frac{\hat{y}}{c_1 \hat{w}}, \qquad x = -1 + \frac{(1-\hat{y}^2)}{2 c_2^2 \hat{w}^2}
\end{equation}  Note that $\hat{y} \in [-1,0)$ and $\hat{w} \to \infty$ if and only if $x \to -1$ and $u \to u_*$. We must restrict the $(\hat{w}, \hat{y})$ plane to region satisfying the analogue of \eqref{condwy} which arises from the restriction $x \leq 1$. 

It is straightfrward to verify that in the asymptotic region $\hat{w} \to \infty$, the asymptotic geometry is given by \eqref{End1} with the replacements $w \to \hat{w}, y \to -\hat{y}$. As before, the $(\hat{y}, \phi)$ coordinates parameterize an open disc with boundary at $\hat{y}=0$. Since the metric is smooth across $\hat{y} =0$, we may continue $\hat{y}$ so that $\hat{y} \in [-1,1]$ with the coordinate singularities at $\hat{y} = \pm 1$ representing fixed points where the Killing field $\partial_\phi$ degenerates smoothly. This extends the metric of the second asymptotic end to a twisted product of a (locally) hyperbolic metric and a round sphere.

\subsection{Static Limit}
It is convenient to consider the subset of `static' solutions with $\alpha=0$ to investigate the behaviour in the asymptotic regions $u \to u_*, x\to \pm 1$ more explicitly. The geometry simplifies considerably and the generators $\partial_\tau, \partial_\phi$ are orthogonal ($g_{\tau\phi} =0$) and hence the local metric \eqref{gext} is diagonal: 
\begin{equation}
    g = \frac{1}{A^2}\left[P(r) \td \tau^2 +  \frac{\td r^2}{P(r)} + r^2 \left(\frac{\td x^2}{(1-x^2)Q(x)} + (1-x^2) Q(x) \td \phi^2\right)\right]
\end{equation} where 
\begin{equation}\begin{aligned}
    P(r) &= (1 + \beta^2 r^2)\left( 1 + \beta^2 m^2 - \frac{2m}{r}\right), \qquad Q(x) = 1 + \beta^2 m^2 x^2, \\
    A(r,x)^2 &= 1 + \beta^2 r^2 - \beta^2 ((1 + \beta^2 m^2)r^2 - 2m r)x^2.
    \end{aligned}
\end{equation} Observe that the normalized Killing fields that generate $2\pi-$periodic flows are
\begin{equation}
    \ell_1 = \left(\frac{1}{1 + \beta^2 m^2}\right)\frac{\partial}{\partial\phi}, \qquad \ell_2 = \left(\frac{4m}{(1 + \beta^2 m^2)^2} \right) \frac{\partial}{\partial\tau}
\end{equation} so that the metric is invariant under the identifications 
\begin{equation}\label{identstatic}
(\psi_1,\psi_2) \sim (\psi_1 + 2\pi, \psi_2), \qquad (\psi_1,\psi_2) \sim (\psi_1, \psi_2 + 2\pi)
\end{equation} where $\psi_1 = (1 + \beta^2 m^2) \phi$ and $\psi_2 = (4m)^{-1}(1 + \beta^2 m^2)^2 \tau$. As before, in the $(\ell_1, \ell_2)$ basis, There are two semi-infinite rods with rod vector $(1,0)$ for $x = \pm 1$ and a finite rod at $r = r_+$ with rod vector $(0,1)$ and
\begin{equation}
    r_+ = \frac{2m}{1 - \beta^2 m^2}. 
\end{equation} and the bound \eqref{mbound} reduces to the requirement $\beta m < 1$. We therefore restrict to the region $r > r_+$ with the coordinate singularity at $r =r_+$ representing a smooth degeneration set of $\ell_2$. Note that the metric function $\Sigma = r^2$ and it is easy to check that $r=0$ is a genuine curvature singularity. 

As discussed above, setting $u = -1/r$ we find that the metric is smooth and bounded across $u =0$ to $u > 0$. The conformal factor takes the explicit form
\begin{equation}
    \frac{1}{A^2} = \frac{u^2}{u^2 - 2\beta^2 m x^2 u + \beta^2(1-x^2) +\beta^4 m^2 x^2}
\end{equation} The denominator is a quadratic in $u$ which has discriminant: 
\begin{equation}
    \mathcal{D} = -4\beta^2(1 - x^2)(1 + \beta^2 m^2 x^2) \leq 0
\end{equation} and so we can only have a real double root when $x= \pm 1$ in which case $u = u_*$ as claimed above. In either root a simple computation gives
\begin{equation}\label{Kretchone}
    R_{abcd}R^{abcd} = 8 \beta^4(1 + \beta^2 m^2)^4.
\end{equation} which suggests that the geometry in this region is well behaved.  First, in the region $u \in (0,u_*)$ we introduce coordinates $(w,y)$ via
\begin{equation}
 w = \left[\left( 1 - \frac{u}{u_*}\right)^2 + \frac{2(1 + \beta^2 m^2)}{m^2\beta^2} (1-x)\right]^{-1/2}, \quad y = \left[1 + \frac{2\beta^2 (1 + \beta^2 m^2)}{(u_* - u)^2} (1-x) \right]^{-1/2}
\end{equation} with inverse
\begin{equation}
    u = u_* \left( 1- \frac{y}{w}\right), \qquad x = 1 - \frac{m^2 \beta^2 (1-y^2)}{2(1 + \beta^2 m^2) w^2}.
\end{equation} Note that $y \in (0,1]$ and $w >0$ with $w\to \infty$ if and only if $x \to 1, u\to u_*$, with $y / w = (u_* - u)/(\beta^2 m) > 0$. In addition, the restriction $x \geq -1$ requires that we restrict to the region of the $(w,y)$ plane with
\begin{equation}
    \frac{1-y^2}{w^2} \leq \frac{4\beta ( 1+ \beta^2 m^2)}{\beta^2 m^2}.
\end{equation} The Jacobian of the transformation $u = u(w,y), x = x(w,y)$ is
\begin{equation}
      \det \left(\frac{\partial(u, x)}{\partial(w,y)} \right) = \frac{\beta^4 m^3}{(1 + \beta^2 m^2) w^4}
\end{equation} which implies that the transformation is locally invertible on an open set, although it fails to be invertible in the limit $w \to \infty$.  The geometry in the asymptotic region $w \to \infty$ simplifies considerably
\begin{equation}\label{staticend1}
\begin{aligned}
    g &=  \frac{(1 + \beta^2 m^2)^2 w^2}{\beta^2 m^2} \td\tau^2\left(1 + O(w^{-1})\right)  + \frac{\td w^2}{\beta^2(1 + \beta^2 m^2)^2 w^2} \left(1 + O(w^{-1})\right) + O(w^{-2}) \td w \td y\\
    & + \frac{\td y^2}{\beta^2(1 + \beta^2 m^2)^2 (1 - y^2)}\left(1 + O(w^{-1})\right) + \frac{(1 - y^2) \td \phi^2}{\beta^2} \left(1 + O(w^{-1})\right).
    \end{aligned}
\end{equation} The asymptotic geometry thus splits into a product. The squared norm of the curvature tensor is given by \eqref{Kretchone} up to $O(w^{-3})$. The $(\tau, w)$ part of the metric is identified with the hyperbolic funnel metric on a punctured disc $(0,\infty) \times S^1$ (i.e. with a compact region containing the cusp at $w =0$ removed). The $(y,\phi)$ part of the metric is defined on an open hemisphere (disc) noting that $y \in (0,1]$. However, the metric and its subleading terms are smooth across $y=0$ and we may extend the metric into the region $-1 < y \leq 0$. The apparent singularity at $y= -1$ is easily seen to represent a smooth degeneration of the Killing field $\ell_1$ (i.e., provided the identifications \eqref{identstatic} are satisfied). The asymptotic metric is then extended to a metric on $(0,\infty) \times S^1 \times S^2$ where the first factor is equipped with a hyperbolic funnel metric.  

To investigate the region near $u \to u_*, x \to -1$, in the region $u \in (0,u_*)$ introduce the coordinate $(\hat{w}, \hat{y})$
\begin{equation}
     \hat{w} = \left[\left( 1 - \frac{u}{u_*}\right)^2 + \frac{2(1 + \beta^2 m^2)}{m^2\beta^2} (1+x)\right]^{-1/2}, \quad \hat{y} = -\left[1 + \frac{2\beta^2 (1 + \beta^2 m^2)}{(u_* - u)^2} (1+x) \right]^{-1/2}
\end{equation}  with inverse
\begin{equation}
    u = u_* \left( 1+ \frac{\hat{y}}{\hat{w}}\right), \qquad x = -1 + \frac{m^2 \beta^2 (1-\hat{y}^2)}{2(1 + \beta^2 m^2) \hat{w}^2}.
\end{equation} Note that $\hat{w} >0$ whereas $\hat{y} \in [-1,0)$. The region $u\to u_*, x\to -1$ corresponds to $\hat{w} \to \infty$.  The requirement $x \leq 1$ restricts the $(\hat{w}, \hat{y})$ coordinates to the region 
\begin{equation}
    \frac{1-\hat{y}^2}{\hat{w}^2} \leq \frac{4\beta ( 1+ \beta^2 m^2)}{\beta^2 m^2}.
\end{equation} The geometry in the asymptotic region $\hat{w} \to \infty$ is
\begin{equation}\label{staticend2}
\begin{aligned}
    g &=  \frac{(1 + \beta^2 m^2)^2 \hat{w}^2}{\beta^2 m^2} \td\tau^2\left(1 + O(\hat{w}^{-1})\right)  + \frac{\td \hat{w}^2}{\beta^2(1 + \beta^2 m^2)^2 \hat{w}^2} \left(1 + O(\hat{w}^{-1})\right) + O(\hat{w}^{-2}) \td \hat{w} \td \hat{y}\\
    & + \frac{\td \hat{y}^2}{\beta^2(1 + \beta^2 m^2)^2 (1 - \hat{y}^2)}\left(1 + O(\hat{w}^{-1})\right) + \frac{(1 - \hat{y}^2) \td \phi^2}{\beta^2} \left(1 + O(\hat{w}^{-1})\right).
    \end{aligned}
\end{equation} To leading order, the metric is isometric to \eqref{staticend1}. We may extend the metric through $\hat{y}=0$ into the region $0 \leq \hat{y} \leq 1$. The apparent singularity at $\hat{y}= 1$ is easily checked to be a smooth degeneration point for $\partial_\phi$ and we can extend the $(\hat{y}, \phi)$ to cover a full $S^2$. The geometry on this asymptotic end extends to $(0,\infty) \times S^1  \times S^2$ where, as above, the first factor is equipped with a hyperbolic funnel metric.  

The two charts $(w,y)$ and $(\hat{w}, \hat{y})$ overlap, with
\begin{equation}
    \hat{w} = \frac{\beta m w}{\left(4(1 + \beta^2 m^2)w^2 + \beta^2 m^2(2y^2-1)\right)^{1/2}}, \qquad \hat{y} = -\frac{\beta m y}{\left(4(1 + \beta^2 m^2)w^2 + \beta^2 m^2(2y^2-1)\right)^{1/2}}.
\end{equation} The Jacobian determinant of this transformation is
\begin{equation}
    \det \left(\frac{\partial(\hat{w}, \hat{y})}{\partial(w,y)} \right)= \frac{\beta^4 m^4}{(4(1 + \beta^2 m^2)w^2 + \beta^2 m^2(2y^2-1))^2}
\end{equation} which vanishes in the limit $w \to \infty$. Thus the $(\hat{w},\hat{y})$ coordinates degenerate in the first asymptotic region. Likewise, we can express $w= w(\hat{w}, \hat{y}), y = y(\hat{w}, \hat{y})$ and compute
\begin{equation}
    \det \left(\frac{\partial(w, y)}{\partial(\hat{w},\hat{y})} \right)= \frac{\beta^4 m^4}{(4(1 + \beta^2 m^2)\hat{w}^2 + \beta^2 m^2(2\hat{y}^2-1))^2}
\end{equation} which vanishes as $\hat{w} \to \infty$. Thus the $(w,y)$ coordinates degenerate and cannot be used to describe the second asymptotic region. This is consistent with the interpretation that the underlying geometry has two separate asymptotic ends.
\section{Discussion}
We have argued that the local family of Riemannian metrics \eqref{gext} extend smoothly to a non compact space with two asymptotic ends, associated to the two distinct regions where the conformal factor diverges. This requires the introducing of new sets of coordinates for which one can smoothly extend the metric across. This behaviour contrasts with typical AF and ALF examples of gravitational instantons, which have a single asymptotic end along with a number of 2 cycle `bolts' in the interior region. The natural example of this is the well known AF Kerr instanton on $\mathbb{R}^2 \times S^2$. The orbit space for this solution in the standard $(r,\theta)$ chart is a semi-infinite strip $[r_+,\infty) \times [0,\pi]$ which clearly has a single asymptotic end with asymptotic boundary $S^1 \times S^2$ (generically, the $S^1$ factor does not have bounded length). In the present case, as a heuristic model, consider the connected sum $M$ of two distinct copies of $\mathbb{R}^2 \times S^2$ in a equivariant manner that preserves the torus action. This can be achieved by removing a ball centred at a corner (fixed point) on the orbit space boundary of each $\mathbb{R}^2 \times S^2$ copy and gluing the result together along the resulting $S^3$ boundaries in such a way as to preserve the overall orientation. The result $M$ will be simply connected and its associated orbit space would have two asymptotic ends, a situation that arises in the Lorentzian setting in the study of extreme black holes \cite{Figueras:2009ci}. It would be interesting to understand the topology of the underlying manifold rigorously.

Finally, the physical motivation for studying gravitational instantons is their role as classical solutions with finite action in Euclidean quantum gravity. In this context they are interpreted as saddle points which dominate the semi-classical gravitational path integral \cite{Gibbons:1976ue}. The thermodynamic properties of the black hole solutions of \cite{Podolsky:2025tle, Ovcharenko:2025cpm} have recently been investigated in \cite{Kubiznak:2026uro}, along with a discussion of their physical properties (recall that they are not asymptotically flat). It may be possible to obtain a similar interpretation of the families of instantons considered here. A natural candidate for an appropriate `background' model is the space \eqref{model}.

\end{document}